\RequirePackage{snapshot}
\pdfoutput=1
\documentclass[a4paper,11pt]{article}

\usepackage{url}

\usepackage{fullpage}

\usepackage{amsmath,amsthm,amssymb}
\usepackage{xcolor}
\usepackage{graphicx}
\usepackage{bbold}
\usepackage{textcomp} 

\usepackage[position=top]{subfig}
\usepackage{authblk}
\renewcommand\Affilfont{\itshape\small}

\title{Efficient Energyminimization in Finite-Difference Micromagnetics: Speeding up Hysteresis Computations}

\author[1]{Claas Abert\thanks{claas.abert@tuwien.ac.at}}
\author[1]{Gregor Wautischer}
\author[1]{Florian Bruckner}
\author[2]{Armin Satz}
\author[1]{Dieter Suess}
\affil[1]{Christian Doppler Laboratory for Advanced Magnetic Sensing and Materials, Institute for Solid State Physics, Vienna University of Technology, Wiedner Hauptstrasse 8-10, 1040 Vienna, Austria}
\affil[2]{Infineon Technologies Austria AG, Siemensstrasse 2, 9500 Villach, Austria}

\begin{document}
\maketitle

\begin{abstract}
  We implement an efficient energy-minimization algorithm for finite-difference micromagnetics that proofs especially useful for the computation of hysteresis loops.
  Compared to results obtained by time integration of the Landau-Lifshitz-Gilbert equation, a speedup of up to two orders of magnitude is gained.
  The method is implemented in a finite-difference code running on CPUs as well as GPUs.
  This setup enables us to compute accurate hysteresis loops of large systems with a reasonable computational effort.
  As a benchmark we solve the \textmu Mag Standard Problem \#1 with a high spatial resolution and compare the results to the solution of the Landau-Lifshitz-Gilbert equation in terms of accuracy and computing time.

\end{abstract}
  
\newpage
\section{Introduction}
For the investigation of ferromagnetic materials and the development of magnetic applications, micromagnetic simulations are an important complement to experiments.
The micromagnetic theory covers both static and dynamical properties of magnetic systems.
While the time evolution of a magnetic system as described by the Landau-Lifshitz-Gilbert equation is of interest for investigations in the high-frequency regime, direct energy minimization is a more suitable approach for the quasi static regime.

A possible application for energy minimization is the computation of hysteresis loops.
These computations can be perceived as dynamical problems with the external field changing over time.
However, a hysteresis loop should resolve the dependence of the magnetization on the external field and the time evolution of the system is not of interest.
Using direct energy minimization can significantly speed up computations.
The retrieval of hysteresis properties is essential for many applications including the design and optimization of permanent magnets, see \cite{skomski_2013}, and GMR sensors, see \cite{treutler_2001}.

In this work we propose an energy minimization algorithm in the framework of finite-difference micromagnetics that was already successfully applied in finite-element micromagnetics \cite{exl_2014}.
We gain a significant speedup of up to two orders of magnitude compared to the solution of the Landau-Lifshitz-Gilbert equation.
The proposed algorithm is implemented in an existing micromagnetic CPU/GPU code. 

This paper is structured as follows.
In section~\ref{sec:algorithm} the minimization algorithm is described in detail.
Section~\ref{sec:implementation} gives a brief overview of the implementation of the algorithm and in section~\ref{sec:experiments} numerical experiments for validation and benchmarking are presented.

\section{Algorithm}\label{sec:algorithm}
The presented method minimizes the total free energy $U(\boldsymbol{m})$ of a magnetic system with the normalized magnetization given by $\boldsymbol{m}$.
The method is adapted from \cite{exl_2014, exl_thesis, goldfarb_2009} and applied to finite-difference micromagnetics.
Consider the continuous version of an unconstrained gradient descent method
\begin{equation}
  \boldsymbol{m}_{k+1} = \boldsymbol{m}_k - \hat{\tau} \left. \frac{\delta U}{\delta \boldsymbol{m}} \right|_{\boldsymbol{m}_k}
  \label{eqn:gradient_descent}
\end{equation}
with $\hat{\tau} > 0$ being the stepsize.
Note that the gradient is replaced by a variational derivative denoted by $\delta / \delta \boldsymbol{m}$ due to the magnetization $\boldsymbol{m}$ being a continuous field.
In order to avoid violation of the micromagnetic constraint $|\boldsymbol{m}| = 1$, the search direction $\boldsymbol{v} = \delta U / \delta \boldsymbol{m}$ of the gradient method is projected onto the tangent plane $\mathcal{T}_m$ of the magnetization, defined by
\begin{equation}
  \mathcal{T}_m = \{ \boldsymbol{x}: \boldsymbol{x} \cdot \boldsymbol{m} = 0 \}.
\end{equation}
The projected direction $\boldsymbol{v}_\text{p}$ is obtained by a Gram-Schmidt like procedure, resulting in
\begin{align}
  \boldsymbol{v}_\text{p}
  &= \frac{\delta U}{\delta \boldsymbol{m}} - \left( \frac{\delta U}{\delta \boldsymbol{m}} \cdot \boldsymbol{m} \right) \boldsymbol{m} \\
  &= \boldsymbol{m} \times \left( \boldsymbol{m} \times \frac{\delta U}{\delta \boldsymbol{m}} \right).
  \label{eqn:projected_direction}
\end{align}
The effective field is given by
\begin{equation}
  \boldsymbol{H}_\text{eff} =
  - \frac{1}{\mu_0 M_\text{s}} \frac{\delta U}{\delta \boldsymbol{m}}
  \label{eqn:effective_field}
\end{equation}
where $\mu_0$ is the vacuum permeability and $M_\text{s}$ is the saturation magnetization.
Inserting this definition into \eqref{eqn:projected_direction}, the constrained gradient method can be written as
\begin{align}
  \boldsymbol{m}_{k+1} = \boldsymbol{m}_k + \tau \boldsymbol{m}_k \times \big( \boldsymbol{m}_k \times \boldsymbol{H}_\text{eff}(\boldsymbol{m}_k) \big)
  \label{eqn:projected_gradient_descent}
\end{align}
where $\tau$ is a scaled version of $\hat{\tau}$.
Although the search direction is confined to the tangent space $\mathcal{T}_m$, this algorithm still violates the micromagnetic constraint for finite stepsize $\tau$.
This deficiency is overcome by application of a midpoint scheme
\begin{equation}
  \boldsymbol{m}_{k+1} = \boldsymbol{m}_k + \tau \frac{\boldsymbol{m}_k + \boldsymbol{m}_{k+1}}{2} \times \big( \boldsymbol{m}_k \times \boldsymbol{H}_\text{eff}(\boldsymbol{m}_k) \big).
  \label{eqn:implicit_gradient_descent}
\end{equation}
This scheme preserves the modulus of $\boldsymbol{m}$, i.e. $|\boldsymbol{m}_{k+1}| = |\boldsymbol{m}_k|$, which can be shown by multiplication with $(\boldsymbol{m}_k + \boldsymbol{m}_{k+1})$.
Note that the midpoint rule is only applied to a single $\boldsymbol{m}_k$-term on the right-hand side, keeping the scheme both linear and local in $\boldsymbol{m}_{k+1}$.
Hence \eqref{eqn:implicit_gradient_descent} can be analytically solved for $\boldsymbol{m}_{k+1}$, see \cite{goldfarb_2009}.
The stepsize $\tau$ is chosen according to the Barzilai-Borwein rule \cite{barzilai_1988} as proposed in \cite{exl_2014, exl_thesis}.
Consider the following auxiliary fields
\begin{align}
  \boldsymbol{g}_k &= \boldsymbol{m}_k \times \big( \boldsymbol{m}_k \times \boldsymbol{H}_\text{eff}(\boldsymbol{m}) \big) \\
  \boldsymbol{s}_{k-1} &= \boldsymbol{m}_k - \boldsymbol{m}_{k-1} \\
  \boldsymbol{y}_{k-1} &= \boldsymbol{g}_k - \boldsymbol{g}_{k-1}.
\end{align}
The timestep for the discrete problem is obtained by taking \eqref{eqn:implicit_gradient_descent} as a quasi-Newton method which yields the following approximate of the Hessian $H$
\begin{equation}
  H = \tau^{-1} \mathbb{1}
\end{equation}
where $\mathbb{1}$ is the identity matrix.
The corresponding secant equation reads $H \boldsymbol{s}_{k-1} = \boldsymbol{m}_{k-1}$.
By projecting the secant equation onto $\boldsymbol{s}_{k-1}$ and $\boldsymbol{m}_{k-1}$ respectively, the following two different solutions for the timestep $\tau_k$ are obtained
\begin{equation}
  \tau^1_k = \frac{\sum_i \boldsymbol{s}^i_{k-1} \cdot \boldsymbol{s}^i_{k-1}}
                  {\sum_i \boldsymbol{s}^i_{k-1} \cdot \boldsymbol{y}^i_{k-1}}
  \quad , \quad
  \tau^2_k = \frac{\sum_i \boldsymbol{s}^i_{k-1} \cdot \boldsymbol{y}^i_{k-1}}
                  {\sum_i \boldsymbol{y}^i_{k-1} \cdot \boldsymbol{y}^i_{k-1}}
  \label{eqn:stepsize}
\end{equation}
where the superscript $i$ denotes the cell number of the discretized field.
We follow the advice given in \cite{goldfarb_2009} and use $\tau^1_k$ and $\tau^2_k$ in an alternating fashion.
A more elaborate switch is proposed in \cite{exl_2014, exl_thesis}.

As stop condition we require the supremum norm of the angular change of the magnetization $\boldsymbol{m}$ divided by the stepsize $\tau$ to be below a certain threshold.
Since \eqref{eqn:stepsize} cannot be used for the computation of the first stepsize, we start with a reasonably small guess for $\tau$.

\section{Implementation}\label{sec:implementation}
The presented algorithm is implemented in the finite-difference code MicroMagnum \cite{micromagnum}.
MicroMagnum uses regular cuboid grids for spatial discretization.
The demagnetization field is computed with an FFT accelerated convolution.
MicroMagnum runs on CPU as well as GPU.

The code was originally built to solve the Landau-Lifshitz-Gilbert equation by numerical integration.
Like other finite-difference codes \cite{oommf,vansteenkiste_2011} it uses explicit Runge-Kutta methods for this task.
Implementation of the presented minimization algorithm is as easy as replacing the Runge-Kutta integration step by \eqref{eqn:implicit_gradient_descent}.
Large parts of the code, especially the effective field contributions, can be reused as is.

\section{Numerical Experiments}\label{sec:experiments}
A demanding benchmark for the computation of a magnetic hysteresis loop is the Standard Problem \#1 as proposed by the \textmu Mag group \cite{mumag1}.
A thin film of size $1000\times2000\times20\,\text{nm}$ with the following material parameters, similar to thoses of Permalloy, is considered
\begin{align}
  M_\text{s} &= 8 \cdot 10^{5} \text{A/m}\\
  A &= 1.3 \cdot 10^{-11} \text{J/m}\\
  K &= 5.0 \cdot 10^{2} \text{J/m}^3
\end{align}
where $M_\text{s}$ is the saturation magnetization, $A$ is the exchange constant and $K$ is the anisotropy constant for a uniaxial anisotropy with the easy axis parallel to the long edge of the sample.
Two hysteresis loops are computed with the external field aligned in the direction of the short edge and the long edge respectively.
The problem definition requires an ``appoximately parallel'' alignment of the field and at the same time it is stated that field-deviations as small as $1^{\circ}$ may significantly change the outcome.
Here we use an in-plane tilting angle of $1^{\circ}$ which is also the choice of many submissions published on the \textmu Mag site.

Even with the same choice of the tilting angle, the published solutions differ significantly.
This is caused by the complexity of the problem on the one hand, because the size of the thin film results in the creation of complicated domain structures.
On the other hand the submitted solutions were computed more than 15 years ago and the computing resources were very limited back then.
This resulted in a comparatively coarse spatial discretisation of the problem which may have led to an inexact description of the involved domain structures.

However, even with today's computing power and mature micromagnetic codes the solution of the Standard Problem \#1 turns out to be nontrivial.
The application of the presented energy-minimizing algorithm enables the computation of a solution with reasonable computational effort.
The solution converges in the sense, that further refinement of the mesh does not change the simulation outcome significantly.

\begin{figure}
  \centering
  \subfloat[]{
    \includegraphics{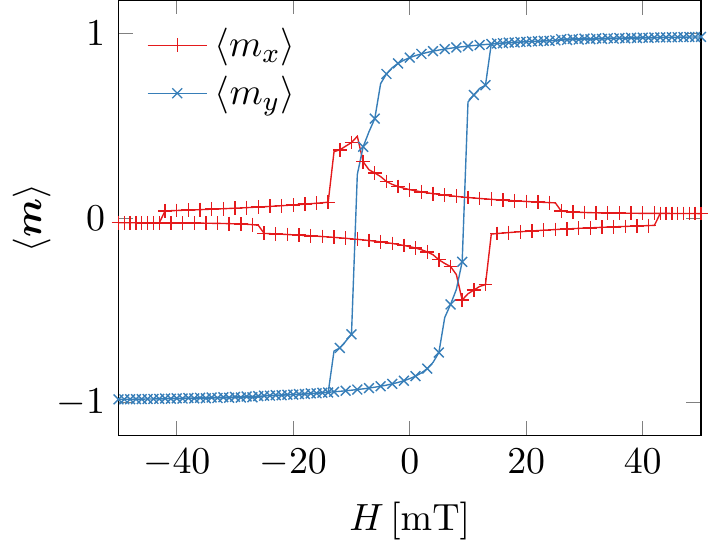}
    \label{fig:sp1_hyst_long}
  }
  \hspace{0.5cm}
  \subfloat[]{
    \includegraphics{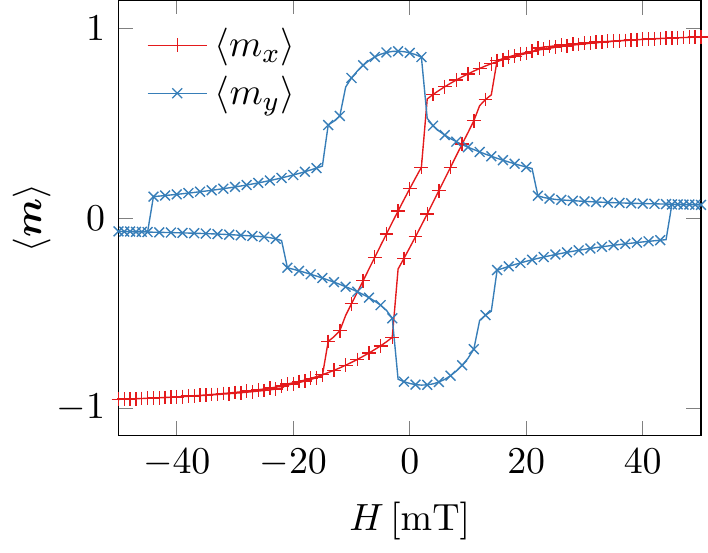}
    \label{fig:sp1_hyst_short}
  }
  \caption{
    Hysteresis curves for the Standard Problem \#1.
    The spatially averaged components of the magnetization $\langle \boldsymbol{m} \rangle$ are plotted against the external field $H$.
    For better readability every second sample point of the external field is omitted.
    (a) External field aligned in direction of the long edge.
    (b) External field aligned in direction of the short edge.
  }
  \label{fig:sp1_hyst}
\end{figure}
Figure~\ref{fig:sp1_hyst} shows the hysteresis curves of the averaged magnetization components.
The external field is varied from -50\,mT to 50\,mT in steps of 1\,mT.
For each applied external field, the minimizing algorithm is run once with the previous magnetization configuration as starting point.
The presented results were computed with a simulation cell-size of $10\times10\times10\,\text{nm}$.
Further refinement of the mesh to $5\times5\times5\,\text{nm}$ does not change the result, see fig.~\ref{fig:sp1_comp}\,(c). 
The same applies to the step size of the field.
Reducing the step size from $1\,\text{mT}$ to $0.5\,\text{mT}$ does not change the outcome of the hysteresis computation.

\begin{figure}
  \centering
  \subfloat[]{
    \includegraphics{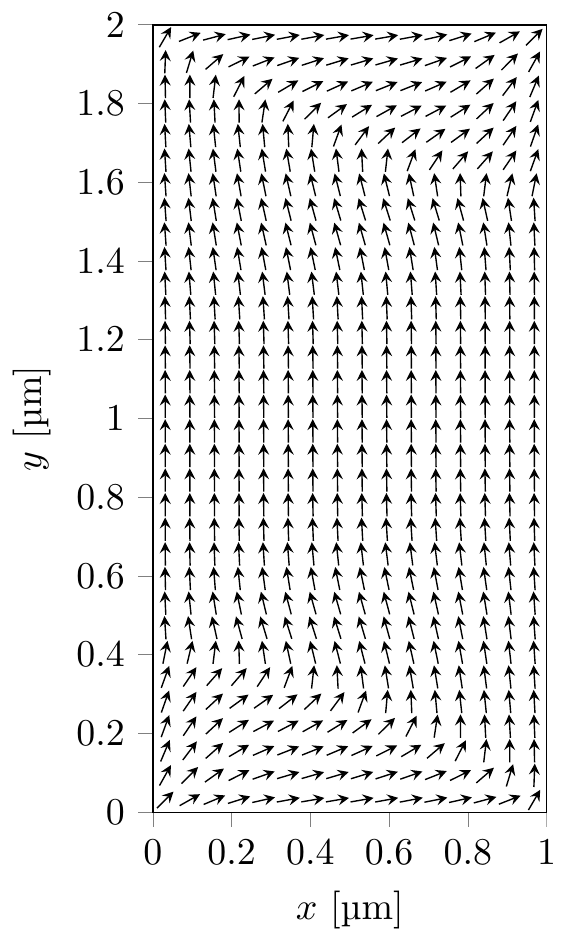}
  }
  \hspace{1cm}
  \subfloat[]{
    \includegraphics{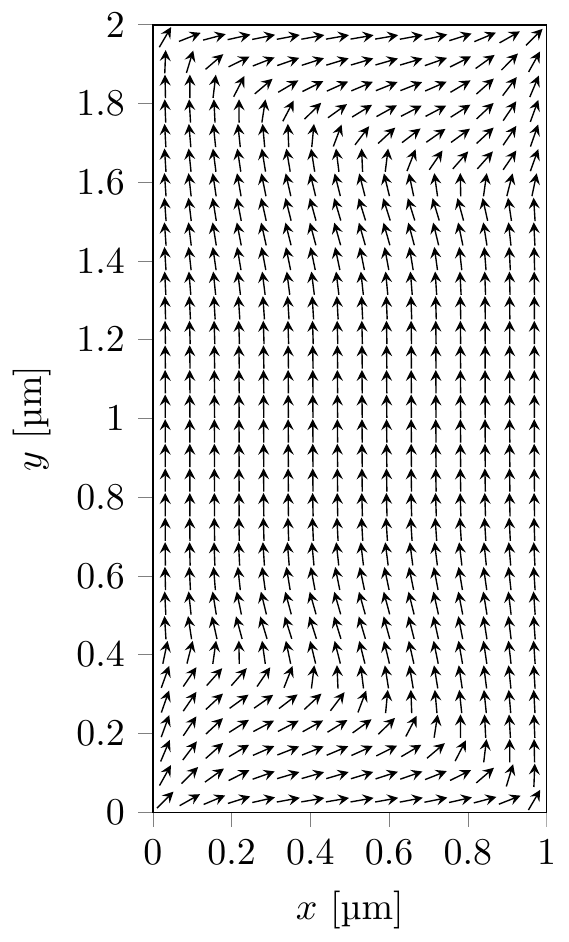}
  }
  \caption{
    Magnetization configuration in the middle-$xy$-plane at remanence.
    For better readability the results are resampled with a spline interpolation.
    (a) Initial external field aligned in direction of the long edge.
    (b) Initial external field aligned in direction of the short edge.
  }
  \label{fig:sp1_m}       
\end{figure}
As required for the submission of the Standard Problem \#1 results, Fig.~\ref{fig:sp1_m} shows the magnetization $\boldsymbol{m}$ at remanence.
As opposed to many submissions to the \textmu Mag site, the remanence magnetization configuration does not depend on the direction of the external field.

\section{Comparison}
The presented method is benchmarked against three alternative approaches for the hysteresis computation.
For the first approach, the magnetization dynamics of the system are calculated by numerical integration of the Landau-Lifshitz-Gilbert equation.
The external field is linearly increased over a relatively large period of time (LLG ramp).
For the remaining two approaches the external field is changed stepwise like for the minimizer algorithm.
Integration of the Landau-Lifshitz-Gilbert equation is then used to find the new energy minimum of the system.
To speed up convergence a high damping of $\alpha = 1$ is chosen for the first method (LLG alpha=1).
For the second method, the precession term of the Landau-Lifshitz-Gilbert equation is omitted completely (LLG no precess).

\begin{figure}
  \centering
  \subfloat[]{
    \includegraphics{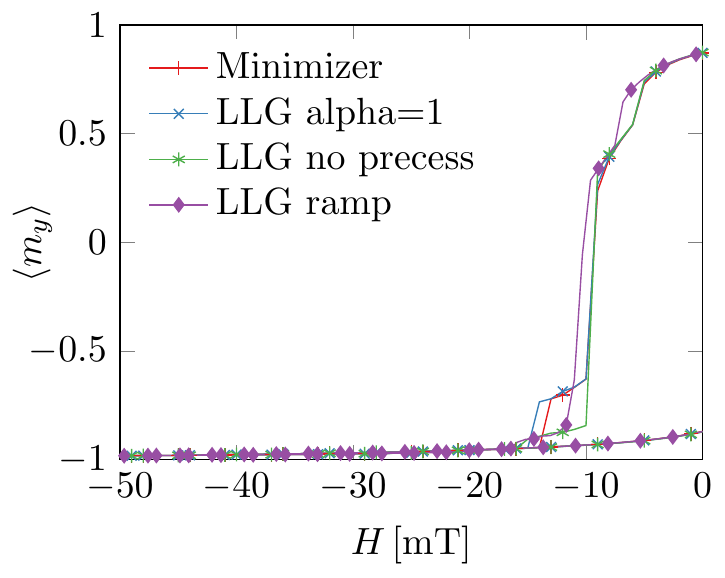}
    \label{fig:sp1_comp_1}
  }
  \subfloat[]{
    \includegraphics{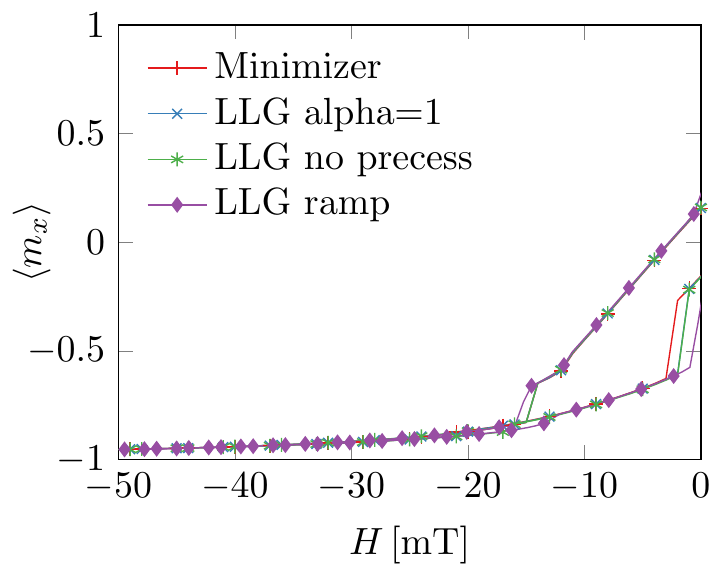}
    \label{fig:sp1_comp_2}
  }
  \\
  \subfloat[]{
    \includegraphics{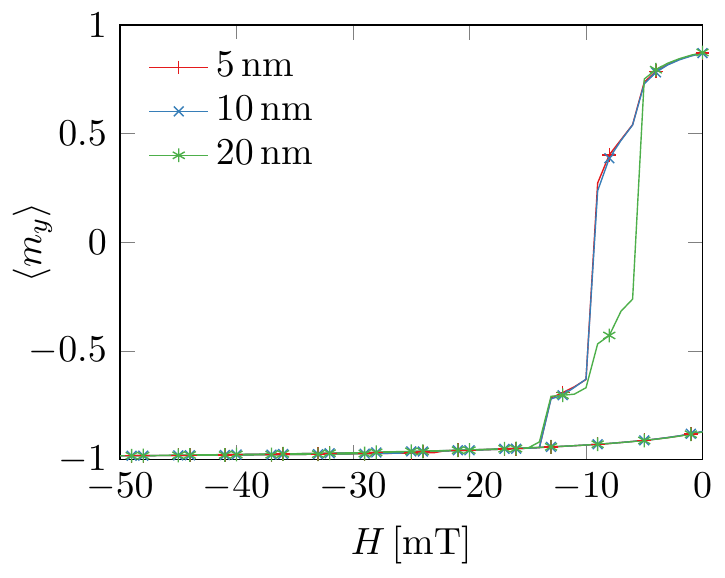}
    \label{fig:sp1_comp_3}
  }
  \subfloat[]{
    \includegraphics{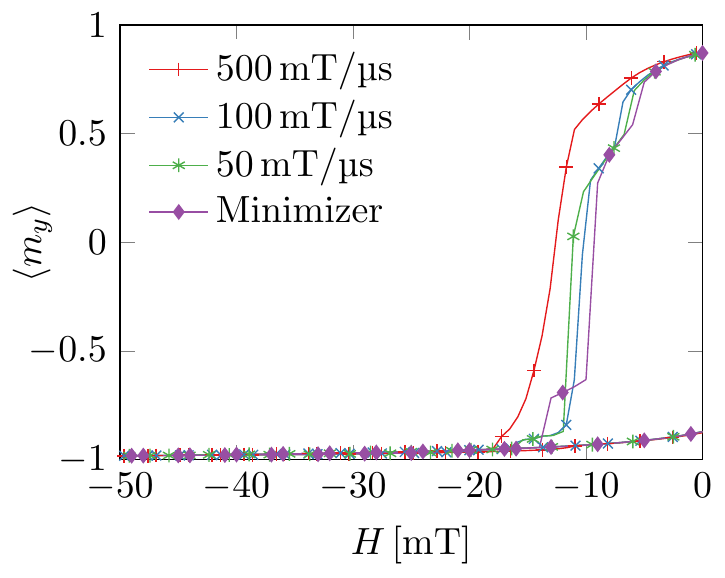}
    \label{fig:sp1_comp_4}
  }

  \caption{
    Comparison of different methods for the hysteresis computation.
    For better readability the symmetric positive fraction of the hysteresis loop is omitted.
    (a) External field aligned in direction of the long edge.
    (b) External field aligned in direction of the short edge.
    (c) Minimizer algorithm for different simulation-cell sizes.
    (d) LLG ramp algorithm for different rise times compared to the solution of the minimizer algorithm.
  }
  \label{fig:sp1_comp}
\end{figure}
Figure~\ref{fig:sp1_comp}\,(a) and (b) show a comparison of the hysteresis loops computed with the different methods.
The simulation parameters are chosen such that the reference solution is essentially reproduced with a computational effort that is as low as possible.
These parameters include the step size of the time integration.
Since we use an adaptive Runge-Kutta scheme for integration, the step size can be increased by increasing the error bounds.
For the stepped methods (LLG alpha=1 and LLG no precess) another parameter is the stop condition that is used for each hysteresis step.
For the continuous method (LLG ramp) the most important parameter is the rise time of the field.
Figure~\ref{fig:sp1_comp}\,(d) shows the hysteresis curve for different rise times in comparison to the reference solution.
In order to obtain viable results with the LLG ramp method, the rise rate should not exceed 100\,mT/\textmu s, which yields a total simulation time of 2\,\textmu s for a complete hysteresis loop.
However, even with this comparably large effort the results are not accurate as shown in fig.~\ref{fig:sp1_comp}\,(d).

\begin{table}
  \centering
  \begin{tabular}{l|r|r}
                    & \multicolumn{2}{c}{\textbf{RHS Evaluations}} \\ 
    \textbf{Method} & Long Edge & Short Edge \\ \hline
    Minimizer       &     64814 &      81454 \\
    LLG no precess  &    835707 &     909627 \\
    LLG alpha=1     &   1387542 &    1018417 \\
    LLG ramp        &   2008265 &    2002229
  \end{tabular}
  \caption{
    Comparison of different methods for the hysteresis computation in terms of right-hand-side (RHS) evaluations.
    Results for the field aligned parallel to the long and the short axis for the \textmu Mag Standard Problem \#1.
  }
  \label{tab:bench}
\end{table}
In order to compare the computational effort required for the hystereris computation, the number of right-hand-side evaluations for a complete hysteresis loop is counted for the different methods.
The results are summarized in tab.~\ref{tab:bench}.
For all methods, the evaluation of the right-hand-side basically requires the computation of the effective field for a given magnetization configuration.
Furthermore all methods have in common, that the calculation of the effective field dominates the overall computational cost.
Thus the simulation time can be assumed to scale linearly with this quantity and it is considered a good measure for performance considerations.

The minimizer algorithm outperforms the alternative algorithms by at least a factor of 10.
Note that the LLG ramp algorithm is by far the slowest while also delivering comparably inaccurate results.
However, improving the accuracy by lowering the field rise rate would further increase the number of right-hand-side evaluations.

\section{Conclusion}
The proposed minimizer algorithm significantly speeds up finite-difference hysteresis computations.
Compared to different flavours of Landau-Lifshitz-Gilbert methods, the proposed method offers a speedup of at least a factor of 10.
The results of the minimizer method converge both in terms of spatial discretization as well as hysteresis step size.
The implementation of the method in existing finite-difference codes is straight forward since large parts of the code can be reused.
We implement the method within the GPU code MicroMagnum \cite{micromagnum}, which enables us to compute hysteresis loops of large systems in reasonable time.

\section*{Acknowledgements}
The authors want to thank Lukas Exl and Thomas Schrefl for valuable advice.
Financial support by
the Austrian Federal Ministry of Economy, Family and Youth and the National Foundation for Research, Technology and Development
is gratefully acknowledged.

\bibliographystyle{ieeetr}
\bibliography{refs}
\end{document}